\documentclass[12pt]{iopart}

\usepackage{graphicx}
\usepackage{bm}

\newcommand{\ket}[1]{\left\vert#1\right\rangle}
\newcommand{\s}{\uparrow}
\newcommand{\g}{\downarrow}

\begin{document}

\title{Electron Fabry-Perot interferometer with two entangled magnetic impurities}

\author{Francesco Ciccarello}

\address{CNISM and Dipartimento di Fisica e
Tecnologie Relative dell'Universit\`{a} degli Studi di Palermo,
Viale delle Scienze, Edificio 18, I-90128 Palermo, Italy}
\ead{ciccarello@difter.unipa.it}

\author{G. Massimo Palma}

\address{ NEST- INFM (CNR) \& Dipartimento di Scienze Fisiche ed Astronomiche dell'Universit\`{a} degli Studi di Palermo, Via Archirafi
36, I-90123 Palermo, Italy}
\ead{massimo.palma@fisica.unipa.it}
\author{Michelangelo Zarcone}

\address{CNISM and Dipartimento di Fisica e
Tecnologie Relative dell'Universit\`{a} degli Studi di Palermo,
Viale delle Scienze, Edificio 18, I-90128 Palermo, Italy}

\author{Yasser Omar}

\address{SQIG, Instituto de Telecomunica\c{c}\~oes, P-1049-001 Lisbon and
CEMAPRE, ISEG, Technical University of Lisbon, P-1200-781 Lisbon,
Portugal}

\author{Vitor R. Vieira}

\address{CFIF and Department of Physics, Instituto Superior
T\'{e}cnico, Av. Rovisco Pais, 1049-001 Lisbon, Portugal}

\begin{abstract}
We consider a one-dimensional (1D) wire  along which single
conduction electrons can propagate in the presence of two spin-1/2
magnetic impurities. The electron may be scattered by each impurity
via a contact-exchange interaction and thus a spin-flip generally
occurs at each scattering event. Adopting a quantum waveguide theory
approach, we derive the stationary states of the system at all
orders in the electron-impurity exchange coupling constant. This
allows us to investigate electron transmission for arbitrary initial
states of the two impurity spins. We show that for suitable electron
wave vectors, the triplet and singlet maximally entangled spin
states of the impurities can respectively largely inhibit the
electron transport or make the wire completely transparent for any
electron spin state. In the latter case, a resonance condition can
always be found, representing an anomalous behaviour compared to
typical decoherence induced by magnetic impurities. We provide an
explanation for these phenomena in terms of the Hamiltonian
symmetries. Finally, a scheme to generate maximally entangled spin
states of the two impurities via electron scattering is proposed.
\end{abstract}


%Uncomment for PACS numbers title message
%\pacs{00.00, 20.00, 42.10}
% Keywords required only for MST, PB, PMB, PM, JOA, JOB?
%\vspace{2pc}
%\noindent{\it Keywords}: Article preparation, IOP journals
% Uncomment for Submitted to journal title message
%\submitto{\JPA}
% Comment out if separate title page not required
\maketitle

\section{Introduction}

The remarkable recent progress in the fabrication techniques of
nanometric semiconductor structures has stimulated a rapid
development of the emerging field of mesoscopic physics \cite{mitin,
davies, datta}. In particular, the fabrication of devices of a size
shorter than the electron coherence length has motivated the study
of systems where the conduction electrons exhibit a fully quantum
mechanical behaviour. Such systems are the electron analogue of
optical devices. For instance, a multichannel quantum wire can be
regarded as the electron counterpart of an electromagnetic waveguide
\cite{datta}.

On the other hand, there is an active research on the coherent
dynamics of the electron spin in mesoscopic systems \cite{ssqc} due
to its potential applications to the control of electron transport
in so-called ``spintronic" devices \cite{dattadas}, as well as in
the implementations of quantum information processing devices
\cite{lossdiv}. Such interest is justified by the long decoherence
times/distances exhibited by electron spin in semiconductors
\cite{ssqc}.

In this paper, we consider a 1D wire with two spin-1/2 magnetic
impurities embedded at fixed positions. The 1D wire could be
realized by a semiconductor quantum wire \cite{davies} or a
single-wall carbon nanotube \cite{nanotube}, while each impurity
could be implemented by means of a single-electron quantum dot
\cite{ssqc}. Conduction electrons entering the wire undergo multiple
scattering by the two impurities before being reflected or
transmitted. At each scattering event the electron and impurity
spins elastically interact via an exchange coupling which can induce
a spin-flip. If the two impurities were static, the present system
would reduce to the electron analogue of a Fabry-Perot (FP)
interferometer with partially silvered mirrors \cite{rossi}, with
the impurities playing the role of the two mirrors. It follows that
our system can be considered as a generalized FP interferometer
where each mirror has a quantum degree of freedom: the spin.

Since scattering with magnetic impurities is a well-known source of
electron decoherence \cite{datta}, one would expect that, when
mirrors with internal degrees of freedom are considered, the typical
resonance condition found in a standard FP interferometer
\cite{rossi} would be modified. The expected loss of electron
coherence is due to the fact that -- contrary to scattering by
static impurities which give rise to well fixed phase-shifts --
scattering by magnetic impurities causes an uncertainty in the phase
shift of the scattered electron \cite{imry,imrybook}. Equivalently,
decoherence can be regarded as a consequence of the unavoidable
entanglement arising between the electron and impurity spin degrees
of freedom \cite{imry, schulman}.

In the present paper, we analyze the case where \emph{two} magnetic
impurities are embedded in the wire. In \cite{ciccarello} we have
investigated how non-local correlations arising when the scatterers
are in an entangled state affect the wire transmittivity. In the
present paper, we will extend our analysis providing all the details
of our approach and expand the discussion of the possible
applications. In particular, we will show that perfect resonance,
i.e.\ perfect transmittivity for suitable electron wave vectors,
appears for the singlet maximally entangled state of the impurity
spins. Remarkably, these resonant wave vectors turn out to be
independent on the electron-impurity coupling constant. Moreover,
when this resonant condition is fulfilled, perfect transmittivity is
obtained for all possible electron spin states. On the contrary, a
large inhibition of electron transmission through the interferometer
is observed for the case of the triplet maximally entangled state.
As we have pointed out in a recent work, the above behaviour of the
singlet and triplet entangled states suggests a novel potential use
of entanglement as a tool to modulate the conductance of a 1D wire
\cite{ciccarello}.

This paper is organized as follows. In section
\ref{system_approach}, we describe in detail the system and the
approach used to derive all the transmission amplitudes needed to
calculate the single electron transmittivity for any arbitrary
initial spin state. In section \ref{one_spin_up}, we discuss the
transmission properties of the interferometer for initial spin
states with only one impurity spin up. Perfect ``transparency"   is
exhibited for the singlet state of the impurities. Section
\ref{conservation} is devoted to the explanation of this phenomenon.
In section \ref{ent_controlled}, we show how our results suggest a
possible use of entanglement of the impurity spins as a tool to
modulate the transmission of the wire. In section
\ref{aligned_spins} we investigate the transmission properties of a
different family of impurity spin states, namely the one for which
both spins are aligned. We show that, in this case, entangled states
exhibit no relevant interference effects. Finally, in section
\ref{ent_gen} we propose a scheme to generate maximally entangled
states of the impurity spins via electron scattering. The form that
an initial spin state must have in order to have perfect
transparency is derived in \ref{appendix}.

\section{System and approach} \label{system_approach}

\begin{figure}
 \includegraphics [scale=0.6]{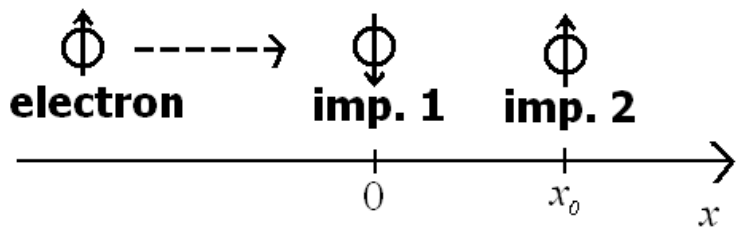}
 \caption{1D wire with two magnetic impurities, labeled 1 and
2, embedded at $x=0$ and $x=x_0$, respectively.}
\label{wire}
\end{figure}

Our system consists of a clean 1D wire into which two spatially
separated, identical spin-1/2 magnetic impurities are embedded. We
assume that single conduction electrons can be injected into the
wire. Due to the presence of an exchange interaction, each
conduction electron undergoes multiple scattering with the
impurities before being transmitted or reflected. Let us also assume
that the electron spin state can be prepared at the input of the
wire and measured at its output (this could be achieved through
ferromagnetic contacts at the source and drain of the wire
\cite{ssqc}). To be more specific, consider a 1D wire along the
$\hat{x}$ direction with the two magnetic impurities, labeled 1 and
2, embedded at $x=0$ and $x=x_0$, respectively, as illustrated in
figure \ref{wire}.
Assuming that the conduction electrons are injected one at a time
(this allows us to neglect many-body effects) and that they can
occupy only the lowest subband, the Hamiltonian can be written as
\begin{equation} \label{H}
H=\frac{p^{2}}{2m^*}+ J \, \mbox{\boldmath$\sigma$}\cdot
\mathbf{S}_{1} \,\delta(x)+ J \, \mbox{\boldmath$\sigma$} \cdot
\mathbf{S}_{2}\,\delta(x-x_{0})
\end{equation}
where $p=-i \hbar \nabla$, $m^*$ and $\bm{\sigma}$  are the electron
momentum operator, effective mass and spin-1/2 operator
respectively, $\mathbf{S}_{i}$ ($i=1,2$) is the spin-1/2 operator of
the $i$-th impurity and  $J$ is the exchange spin-spin coupling
constant between the electron and each impurity. All the spin
operators are in units of $\hbar$. Since the electron-impurity
collisions are elastic, the energy eigenvalues are simply
$E=\hbar^{2} k^2/2m^*$ ($k>0$) where $k$ is a good quantum number.
As the total spin Hilbert space is 8D and considering left-incident
electrons, it turns out that to each value of $k$ there corresponds
an 8-fold degenerate energy level. Let
$\mathbf{S}=\bm{\sigma}+\mathbf{S}_{1}+\mathbf{S}_{2}$ be the total
spin of the system. Since $\mathbf{S^2}$ and $S_z$, with quantum
numbers $s$ and $m_s$, respectively, are constants of motion, $H$
can be block diagonalized, each block corresponding to an eigenspace
of fixed $s$ (for three spins 1/2, the possible values of $s$ are
$1/2, 3/2$) and $m_s=-s,...,s$ (from now on the subscript $s$ in
$m_s$ will be omitted). Let us rewrite equation (\ref{H}) in the
form
\begin{equation} \label{H2}
H=\frac{p^{2}}{2m^*}+\frac{J}{2}
\left(\mathbf{S}_{e1}^{2}-\frac{3}{2}\right)
\delta(x)+\frac{J}{2}\left(\mathbf{S}_{e2}^{2}-\frac{3}{2}\right)
\delta(x-x_{0})
\end{equation}
where $\mathbf{S}_{ei}=\bm{\sigma}+\mathbf{S}_{i}$ ($i=1,2$) is the
total spin of the electron and the $i$-th impurity. Note that in
general $\mathbf{S}_{e1}^{2}$ and $\mathbf{S}_{e2}^{2}$ do not
commute.

Here we choose as spin space basis the states $\ket{s_{e2}; s, m}$,
common eigenstates of $\mathbf{S}_{e2}^{2}$, $\mathbf{S}^{2}$ and
$S_z$ \cite{nota1}, to express, for a fixed $k$, each of the eight
stationary states of the system as an 8D column. Since
$\mathbf{S}_{e1}^{2}$ and $\mathbf{S}_{e2}^{2}$ do not commute, the
latter in general is not a constant of motion and thus $s_{e2}$ in
general is not a good quantum number.

%Therefore, as it will be more clear later, we denote each stationary
%state of the system as $\ket{\Psi_{k, s'_{e2};s,m}}$, where
%$s'_{e2}$ ($s=1/2 \Rightarrow s'_{e2}=0,1$; $s=3/2\Rightarrow
%s'_{e2}=1$) is a labeling index which generally differs from
%$s_{e2}$. $s'_{e2}$ stand for the value of $s_{e2}$ of the incoming
%part of the stationary state.

To determine the transmission properties of the interferometer for a given arbitrary initial spin state, we have to calculate the
transmission probability amplitudes $t_{s_{e2}}^{(s'_{e2};s)}$ that an electron prepared in the incoming state
$\ket{k}\ket{s'_{e2}; s, m}$ is transmitted in the state $\ket{k}\ket{s_{e2}; s, m}$. The calculation of
$t_{s_{e2}}^{(s'_{e2};s)}$ requires the exact stationary states of the system to be derived. To do this, we properly generalize
the quantum waveguide theory approach of reference \cite{amjphys} for an electron scattering with a single magnetic impurity to
the case of two impurities. Note that due to the form of $H$ (see equation \ref{H2}) coefficients $t_{s_{e2}}^{(s'_{e2};s)}$ do
not depend on $m$, as it will be more clear in the following. We first consider the subspace $s=3/2$ and then the subspace
$s=1/2$.

\subsection{Subspace $s=3/2$} \label{subspace_s32}
In this 4D subspace ($m=-3/2,-1/2,1/2,3/2$), both $s_{e1}$ and $s_{e2}$ can assume only the value 1. It follows that in this
subspace $\mathbf{S}_{e1}^{2}$ and $\mathbf{S}_{e2}^{2}$ effectively commute. The states $\ket{s_{e2}; s, m} = \ket{1; 3/2, m}$
are thus also eigenstates of $\mathbf{S}_{e1}^{2}$ and the effective electron-impurities potential $V$ in equation (\ref{H2})
reduces to
\begin{equation} \label{V32}
V=\frac{J}{4} \, \delta(x)+\frac{J}{4}\, \delta(x-x_{0})
\end{equation}
Note that the two impurities behave as if they were static and the scattering between electron and impurities cannot flip the
spins. The four stationary states take therefore the simple product form
\begin{equation}
\ket{\Psi_{k,1;3/2,m}} = \ket{\phi_{k}}\ket{1; 3/2, m}
\end{equation}
where the second index in the left-hand side stand for $s_{e2}=1$
and where $\ket{\phi_{k}}$ describes the electron orbital degrees of
freedom. To determine the wave function $\phi_{k}(x)$, we split the
$\hat{x}$ axis into the three domains $x<0$, $0<x<x_{0}$ and
$x>x_{0}$ labeled I, II and III, respectively. Solving the
Schr\"{o}dinger equation, the wave function $\phi_{k,i}(x)$ in each
domain $i=I, II, III$ is readily written as
\begin{eqnarray}
\phi_{k,I}(x)&=&A_{I}e^{ikx}+B_{I}e^{-ikx}\
\label{wavefunction1}\\
\phi_{k,II}(x)&=&A_{II}e^{ikx}+B_{II}e^{-ikx}
\label{wavefunction2}\\
\phi_{k,III}(x)&=& t_{1}^{(1;3/2)}e^{ikx}
\label{wavefunction3}
\end{eqnarray}
Setting $A_{I}$ to unity, the other four coefficients appearing in
equations (\ref{wavefunction1}) -- (\ref{wavefunction3}) can be
found by requiring the wave function to be continuous at the two
boundaries $x=0$ and $x=x_0$ and its derivative $\phi'_{k}(x)$ to
exhibit a jump at the same points according to
\begin{eqnarray}
\phi'^{+}_{k}(0)-\phi'^{-}_{k}(0)&=&\frac{2m^*}{\hbar^2} \frac{J}{4}\,\phi_{k}(0)\label{jumps32_1}\\
\phi'^{+}_{k}(x_0)-\phi'^{-}_{k}(x_0)&=&\frac{2m^*}{\hbar^2} \frac{J}{4}\,\phi_{k}(x_0) \label{jumps32_2}
\end{eqnarray}
The last conditions are easily found integrating the Schr\"{o}dinger
equation across each impurity (see e.g. \cite{mao}). This yields the
following result for the transmission amplitude $t_{1}^{(1;3/2)}$
\begin{equation}\label{t3/2}
t_{1}^{(1;3/2)}=\frac{64}{64+\pi\rho(E)J\,[16i+(e^{2ikx_0}-1)\rho(E)J]}
\end{equation}
where $\rho(E)=(\sqrt{2m^{*}/E})/\pi\hbar$ is the density of states
per unit length of the wire \cite{mitin, davies}. $t_{1}^{(1;3/2)}$
is thus a function of the two dimensionless parameters $kx_0$ and
$\rho(E)J$. Note that it does not depend on $m$ due to the effective
form (\ref{V32}) of $V$.

\subsection{Subspace $s=1/2$}  \label{subspace_s12}

In this 4D subspace $s_{e1},s_{e2}=0,1$ and thus
$\mathbf{S}_{e1}^{2}$ and $\mathbf{S}_{e2}^{2}$ do not commute and
$s_{e2}$ is not a good quantum number. This is a signature of the
fact that in this space spin-flip may occur. In each of the 2D
$m=-1/2,1/2$ subspaces, the two stationary states are thus of the
form
\begin{equation}  \label{ansatz}
\ket{\Psi_{k,s_{e2}';1/2,m}}=\sum_{s_{e2}=0,1}\ket{\varphi_{k,s_{e2}',s_{e2}}}\ket{s_{e2};
1/2, m}
\end{equation}
where we have used the labeling index $s_{e2}'=0,1$ to indicate that
the incident spin state of (\ref{ansatz}) is $\ket{s_{e2}';1/2,m}$
and where $\ket{\varphi_{k,s_{e2}',s_{e2}}}$ describe the electron
orbital degrees of freedom (from now on we omit the index $k$). Note
that in the case $s=3/2$ discussed in subsection \ref{subspace_s32}
$s'_{e2}$ and $s_{e2}$ coincide.

For a fixed $s'_{e2}=0,1$ the two wave functions
$\varphi_{s'_{e2},0}(x)$ and $\varphi_{s'_{e2},1}(x)$ in each domain
$i=I,II,III$ turn out to take a form analogous to
(\ref{wavefunction1}) -- (\ref{wavefunction3})
\begin{eqnarray}
\varphi_{s'_{e2},s_{e2},I}(x)&=&A_{s_{e2},\,I}^{(s'_{e2})}\,e^{ikx}+B_{s_{e2},\,I}^{(s'_{e2})}\,e^{-ikx}
\label{wavefunction_II_1}\\
\varphi_{s'_{e2},s_{e2},II}(x)&=&A_{s_{e2},\,II}^{(s'_{e2})}\,e^{ikx}+B_{s_{e2},\,II}^{(s'_{e2})}\,e^{-ikx}
\label{wavefunction_II_2}\\
\varphi_{s'_{e2},s_{e2},III}(x)&=&t_{s_{e2}}^{(s'_{e2};1/2)}\,e^{ikx} \label{wavefunction_II_3}
\end{eqnarray}
with $s_{e2}=0,1$. According to the above definition of $s'_{e2}$, the stationary state corresponding to a given $s'_{e2}$ is
obtained by setting $A_{0,\,I}^{(s'_{e2}=0)}=1$, $A_{1,\,I}^{(s'_{e2}=0)}=0$ and $A_{0,\,I}^{(s'_{e2}=1)}=0$,
$A_{1,\,I}^{(s'_{e2}=1)}=1$. In each case, one has to determine the remaining eight coefficients $B_{0,\,I}^{(s'_{e2})},
A_{0,\,II}^{(s'_{e2})}, B_{0\,,II}^{(s'_{e2})}, t_{0}^{(s'_{e2};1/2)}, B_{1,\,I}^{(s'_{e2})}, A_{1,\,II}^{(s'_{e2})},
B_{1\,,II}^{(s'_{e2})}, t_{1}^{(s'_{e2};1/2)}$.
%(the first four defining $\varphi_{s'_{e2},0}(x)$ and the remaining
%$\varphi_{s'_{e2},1}(x)$, respectively).
To do this, we need eight constrains. Four of these are obtained imposing the continuity of both $\varphi_{s'_{e2},0}(x)$ and
$\varphi_{s'_{e2},1}(x)$ at $x=0$ and $x=x_0$. The other four constrains come from appropriate boundary conditions for the
derivatives of $\varphi_{s'_{e2},0}(x)$ and $\varphi_{s'_{e2},1}(x)$ at the impurities' sites. To derive these, we insert the
ansatz (\ref{ansatz}) into the Schr\"{o}dinger equation
%\begin{eqnarray}\label{scrhrodinger1}
%\left[\frac{\hbar^2}{2m^{*}}\frac{d^2}{dx^2}-\frac{J}{2}
%\left(\mathbf{S}_{e1}^{2}-\frac{3}{2}\right)
%\delta(x)+E\right]\left\{\sum_{s_{e2}=0,1}\ket{\varphi_{k,s_{e2}',s_{e2}}}\ket{s_{e2};
%1/2, m}=0 \right\}\\
%\left[\frac{\hbar^2}{2m^{*}}\frac{d^2}{dx^2}-\frac{J}{2}
%\left(\mathbf{S}_{e2}^{2}-\frac{3}{2}\right)
%\delta(x-x_0)+E\right]\left\{\sum_{s_{e2}=0,1}\ket{\varphi_{k,s_{e2}',s_{e2}}}\ket{s_{e2};
%1/2, m}=0\right\} \label{scrhrodinger2}
%\end{eqnarray}
%\begin{eqnarray}\label{scrhrodinger1}
%\nonumber \Biggl[\frac{\hbar^2}{2m^{*}}\frac{d^2}{dx^2}-\frac{J}{2}
%\left(\mathbf{S}_{e1}^{2}-\frac{3}{2}\right) \delta(x)-\frac{J}{2}
%\left(\mathbf{S}_{e2}^{2}-\frac{3}{2}\right)
%\delta(x-x_0)\\
%\,\,\,\,\,\,\,\,\,\,\,\,\,\,\,\,\,\,\,\,\,\,\,\,\,\,\,\,\,\,\,\,\,\,\,\,\,
%\,\,\,\,\,\,+E\Biggl]\left\{\sum_{s_{e2}=0,1}\ket{\varphi_{k,s_{e2}',s_{e2}}}\ket{s_{e2};
%1/2, m}=0\right\} \label{scrhrodinger2}
%\end{eqnarray}
\begin{eqnarray}\label{scrhrodinger}
\fl\Biggl\{\frac{p^2}{2m^{*}}+\frac{J}{2}
\left(\mathbf{S}_{e1}^{2}-\frac{3}{2}\right) \delta(x)+\frac{J}{2}
\left(\mathbf{S}_{e2}^{2}-\frac{3}{2}\right) \delta(x-x_0)-E\Biggl\}
\sum_{s_{e2}=0,1}\ket{\varphi_{s_{e2}',s_{e2}}}\ket{s_{e2}; 1/2, m}=0\nonumber\\
\end{eqnarray}
We now project both sides of equation (\ref{scrhrodinger}) onto
$\ket{0; 1/2, m}$ and $\ket{1; 1/2, m}$. This yields the two
equations
\begin{eqnarray}
\fl\nonumber\left\{
-\frac{\hbar^2}{2m^{*}}\frac{d^2}{dx^2}+\frac{J}{2}\, \left[\langle
0 \left|\mathbf{S}_{e1}^{2}\right|0 \rangle-\frac{3}{2} \right]
\delta(x)-\frac{3J}{4}\, \delta(x-x_0) -\frac{\hbar^2k^2}{2m}
\right\} \ket{\varphi_{s_{e2}',0}} \nonumber \\+\frac{J}{2} \langle
1 \left|\mathbf{S}_{e1}^{2}\right|0 \rangle
\,\delta(x)\ket{\varphi_{s_{e2}',1}}= 0 \label{coupledeqs1} \\
 \fl \nonumber\left\{-\frac{\hbar^2}{2m^{*}}\frac{d^2}{dx^2}+\frac{J}{2}
 \left[ \,\langle 1 \left|\mathbf{S}_{e1}^{2}\right|1 \rangle-\frac{3}{2} \right]
 \delta(x)+\frac{J}{4}\,\delta(x-x_0)-\frac{\hbar^2k^2}{2m}\right\} \ket{\varphi_{s_{e2}',1}}
\nonumber \\+\frac{J}{2} \langle 1 \left|\mathbf{S}_{e1}^{2}\right|0
\rangle \,\delta(x)\ket{\varphi_{s_{e2}',0}}=0\label{coupledeqs2}
\end{eqnarray}
where $\ket{0}$ and $\ket{1}$ stand for $\ket{0; 1/2, m}$ and
$\ket{1; 1/2, m}$, respectively. The matrix elements of
$\mathbf{S}_{e1}^{2}$ appearing in (\ref{coupledeqs1}) --
(\ref{coupledeqs2}) can be computed through a change of the coupling
scheme expressing basis states $\ket{s_{e2}; 1/2, m}$ in terms of
$\ket{s_{e1}; 1/2, m}$ by means of $6j$ coefficients. This yields
\begin{equation} \label{matrix_coeff}
\langle 0 \left|\mathbf{S}_{e1}^{2}\right|0 \rangle=\frac{3}{2}
\,\,\,\,\,\,\,\,\,\,\, \langle 1\left|\mathbf{S}_{e1}^{2}\right|0
\rangle= \langle 0\left|\mathbf{S}_{e1}^{2}\right|1
\rangle=\frac{\sqrt{3}}{2} \,\,\,\,\,\,\,\,\,\, \langle 1
\left|\mathbf{S}_{e1}^{2}\right|1 \rangle=\frac{1}{2}
\end{equation}
Using (\ref{matrix_coeff}) and integrating both equations
(\ref{coupledeqs1}) and (\ref{coupledeqs2}) across $x=0$ and
$x=x_0$, we end with the four equations
\begin{eqnarray}
\Delta \varphi_{s_{e2}',0}'(0)= \frac{2m^*J}{\hbar^{2}} \frac{\sqrt{3}}{4}\,\varphi_{s_{e2}',1}(0)\label{boundcondacc1}\\
\Delta \varphi_{s_{e2}',1}'(0)=-\frac{2m^*J}{\hbar^{2}}\frac{1}{2}\,\varphi_{s_{e2}',1}(0) +  \frac{2m^*J}{\hbar^{2}}\frac{\sqrt{3}}{4} \,\varphi_{s_{e2}',0}(0)\label{boundcondacc2} \\
\Delta \varphi_{s_{e2}',0}'(x_0)=-\frac{2m^*J}{\hbar^{2}}\frac{3}{4}\,\varphi_{s_{e2}',0}(x_0)  \label{boundcondacc3}\\
\Delta
\varphi_{s_{e2}',1}'(x_0)=\frac{2m^*J}{\hbar^{2}}\,\frac{1}{4}\,\varphi_{s_{e2}',1}(x_0)\label{boundcondacc4}
\end{eqnarray}
where $\Delta
\varphi'_{s_{e2}',s_{e2}}(x)=\varphi_{s_{e2}',s_{e2}}^{'+}(x)-\varphi_{s_{e2}',s_{e2}}^{'-}(x)$.

Equations (\ref{boundcondacc1}) -- (\ref{boundcondacc4}) represent
the appropriate boundary conditions for the derivatives of
$\varphi_{s'_{e2},0}(x)$ and $\varphi_{s'_{e2},1}(x)$ at the
impurities' sites. Note that these imply a coupling between
$\ket{\varphi_{s'_{e2},0}}$ and $\ket{\varphi_{s'_{e2},1}}$, as
witnessed by equations (\ref{boundcondacc1}) and
(\ref{boundcondacc2}) and ultimately by the terms proportional to
$\langle 1\left|\mathbf{S}_{e1}^{2}\right|0 \rangle$ appearing in
(\ref{coupledeqs1}) and (\ref{coupledeqs2}). This coupling thus
results from non commutation of $\mathbf{S}_{e1}^{2}$ and
$\mathbf{S}_{e2}^{2}$.

As equations (\ref{boundcondacc1}) -- (\ref{boundcondacc4}) are
added to the matching conditions of $\varphi_{s'_{e2},0}(x)$ and
$\varphi_{s'_{e2},1}(x)$ at $x=0$ and $x=x_0$ a linear system of 8
equations is obtained. Once this is solved for $s'_{e2}=0$ and
$s'_{e2}=1$, the following transmission amplitudes
$t_{s_{e2}}^{(s'_{e2};1/2)}$ are obtained
\begin{eqnarray}\label{t0t1_1}
 t^{(s_{e2}'; 1/2)}_{0}&=&\frac{1}{\delta}\left[-64\,e^{2ikx_0}\pi^2(\rho(E)J)^2\left(2(1-s_{e2}')+\sqrt{3}s_{e2}'\right)\right. \\
 &+ &\left.64(\pi\rho(E)J-8i) \left(2(4i+\pi\rho(E)J)(1-s_{e2}')+\sqrt{3}\pi \rho(E)Js_{e2}'\right)\right]\nonumber \\
\nonumber \\
\label{t0t1_2} t^{(s_{e2}';
1/2)}_{1}&=&\frac{64}{\delta}\left[\sqrt{3}\pi\rho(E)J
\left(-8i+3(e^{2ikx_0}-1)\pi
\rho(E)J\right)(1-s_{e2}')  \right. \\
&+&\left.\,8s_{e2}'(8-3i\pi\rho(E)J) \right] \nonumber
\end{eqnarray}
with
\begin{eqnarray} \label{delta}
\delta &=&\left\{4096+\pi  \rho(E)J \left[-2048 i+\left(e^{2 i k
x_0}-1\right) \pi  \rho(E)J  \right.\right.\nonumber \\
&\times &\left.\left. \left(-128+96 i \pi \rho(E)J +9 \left(e^{2 i k
x_0}-1\right) \pi ^2 (\rho(E)J) ^2\right)\right]\right\}
\end{eqnarray}
As in the case $s=3/2$, the coefficients
$t_{s_{e2}}^{(s'_{e2};1/2)}$ are functions of $kx_0$ and $\rho(E)J$
and, moreover, they again do not depend on $m$ as suggested by the
notation here adopted. The latter circumstance is due to the form
(\ref{H2}) of $H$ and to the fact that $6j$ coefficients -- and thus
matrix elements (\ref{matrix_coeff}) appearing in
(\ref{coupledeqs1}), (\ref{coupledeqs2}) -- do not depend on $m$
(see e.g. \cite{weissbluth}).

As a further signature of non conservation of $\mathbf{S}_{e2}^2$ in
the present subspace note that $t^{(s_{e2}'; 1/2)}_{s_{e2}}\neq 0$
for $s_{e2}\neq s_{e2}'$.

\subsection{Calculation of transmittivity for an arbitrary spin
state} \label{calcul_chi}

It is important to stress again that our calculated transmission
amplitudes $t^{(s_{e2}'; 1/2)}_{s_{e2}}$ are exact at {\em all}
orders in the electron-impurity coupling constant $J$. This follows
from our quantum waveguide theory approach which addresses the
determination of the stationary states through resolution of the
Schr\"{o}dinger equation. This approach is different from the
perturbative one adopted in \cite{yasser} where  only a finite
number of electron multiple reflections between the two impurities
are taken into account performing a few iterations of the Fermi
Golden rule.

The knowledge of all coefficients $t_{s_{e2}}^{(s'_{e2};s)}$
completely describes the transmission properties of our system. Here
we are mainly interested in calculating how an electron with a given
wave vector $k$ and for some initial electron-impurities spin state
$\ket{\chi}$ is transmitted through the wire. Thus assuming to have
the incident wave $\ket{k}\ket{\chi}$, with $\ket{\chi}$ being an
arbitrary spin state, it is straightforward to see that
$\ket{k}\ket{\chi}$ is the incoming part of the stationary state
\begin{equation}\label{sviluppochi}
\ket{\Psi_{k,\chi}}=\sum_{s_{e2}',s,m} \langle s_{e2}';s,m
\ket{\chi} \ket{\Psi_{k,s_{e2}';s,m}}
\end{equation}
where $s_{e2}'=1$ for $s=3/2$ , while $s_{e2}'=0,1$ for $s=1/2$. It
follows that the transmitted part $\ket{\Psi_{k,\chi}}_t$ of
(\ref{sviluppochi}) provides the transmitted state into which
$\ket{k}\ket{\chi}$ evolves after multiple scattering. To calculate
$\ket{\Psi_{k,\chi}}_t$ we simply replace each stationary state
$\ket{\Psi_{k,s_{e2}';s,m}}$ in (\ref{sviluppochi}) with its
transmitted part, express the latter in terms of amplitudes
$t^{(s_{e2}'; s)}_{s_{e2}}$ and rearrange (\ref{sviluppochi}) as a
linear expansion in the basis $\ket{s_{e2};s,m}$. This yields
\begin{equation}\label{sviluppochi_t}
\ket{\Psi_{k,\chi}}_t=\ket{k}  \sum_{s_{e2},s,m}
\gamma_{s_{e2},s,m}(\chi) \ket{s_{e2},s,m} \nonumber \\
\end{equation}
with
\begin{equation} \label{alpha_coeff}
\gamma_{s_{e2},s,m}(\chi) =\sum_{s_{e2}'} t^{(s_{e2}'; s)}_{s_{e2}}
\langle s_{e2}';s,m \ket{\chi}
\end{equation}
Coefficients (\ref{alpha_coeff}) fully describe how an incoming wave
$\ket{k}\ket{\chi}$ is transmitted after scattering. For instance,
the total electron transmittivity $T$ can be calculated as
\begin{equation} \label{T}
T =\sum_{s_{e2},s,m} |\gamma_{s_{e2},s,m}(\chi)|^2
\end{equation}

\section{Transmission properties for one impurity spin up-states} \label{one_spin_up}

In this section, we investigate how the electron transmission is affected by an initial spin state of the two impurity spins
belonging to the family
\begin{equation} \label{family2}
\ket{\Psi (\vartheta , \varphi)} = \cos\vartheta\ket{\s\g} +
e^{i\varphi}\sin\vartheta\ket{\g\s}
\end{equation}
with $\vartheta\in[0,2\pi]$ and $\varphi\in[0,\pi]$. This family
describes all the states with only one impurity spin up, including
both maximally entangled and product states. Following the
calculation scheme illustrated in subsection (\ref{calcul_chi}), the
electron transmittivity $T$ when the injected electron spin state is
$\ket{\s}$ can thus be computed setting
$\ket{\chi}=\ket{\s}\ket{\Psi (\vartheta , \varphi)}$. The behaviour
of $T$ when the impurities are prepared in the product states
$\ket{\s\g}$ ($\vartheta=0$) and $\ket{\g\s}$ ($\vartheta=\pi/2$) is
plotted in figures 2(a) and (b), respectively.
\begin{figure}[htbp]
              {\hbox{ {\includegraphics[scale=0.7]{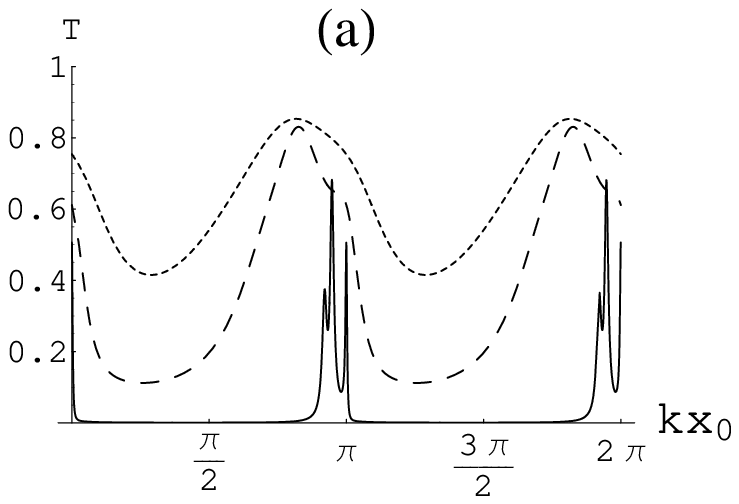}}
{\includegraphics[scale=0.7]{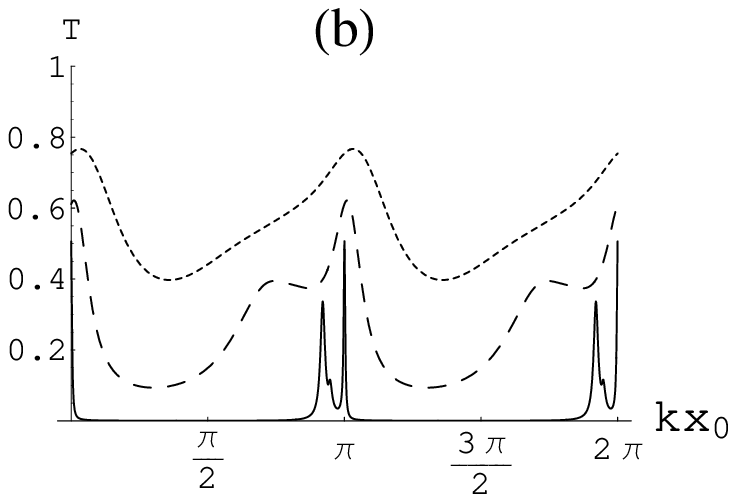}
              }}}
\caption{\footnotesize {Electron transmittivity $T$ as a function of
$kx_{0}$ when the electron is injected in the state $\ket{\s}$ with
the impurities prepared in the state $\ket{\s\g}$ (a) and
$\ket{\g\s}$ (b). Dotted, dashed and solid lines stand for
$\rho(E)J=1,2,10$, respectively.}}
\end{figure}
A behaviour similar to a FP interferometer with partially silvered
mirrors \cite{rossi}, with equally spaced maxima of transmittivity,
is exhibited. In figure 2(a) principal maxima occur around a value
of $kx_{0}\neq n\pi$ ($n$ integer) which tends to $kx_{0} =n \pi$
for increasing values of $\rho(E)J$, while in figure 2(b) they occur
at $kx_{0} =n \pi$. As $\rho(E)J$ is increased, maxima become lower
and sharper. Remarkably, in both cases the electron and impurities
spin state is changed after the scattering (as resulting from the
calculated coefficients $\gamma_{s_{e2},s,m}(\chi)$) and the
electron undergoes a loss of coherence, since we always have $T<1$
\cite{datta,imry,joshi}. The above product impurity spins states
thus lead to the typical decoherent behaviour encountered with
magnetic impurities which avoids a perfect resonance condition $T=1$
to occur (see the Introduction).

We now consider maximally entangled spin states belonging to the
family (\ref{family2}) for $\vartheta = \pi/4$. Let us start with
the triplet state $\ket{\Psi^{+}} = (\ket{\s\g}
+\ket{\g\s})/\sqrt{2}$ (see figure 3(a)). A behaviour  similar to
the case of figure 2 is exhibited. Again the transmitted spin state
differs from the incident one, this indicating occurrence of
spin-flip. In particular, when $kx_{0} =n \pi$, the transmitted
state turns out to be a linear combination of
$\ket{\s}\ket{\Psi^{+}}$ and $\ket{\g}\ket{\s\s}$.
\begin{figure}[htbp]
              {\hbox{ {\includegraphics[scale=0.7]{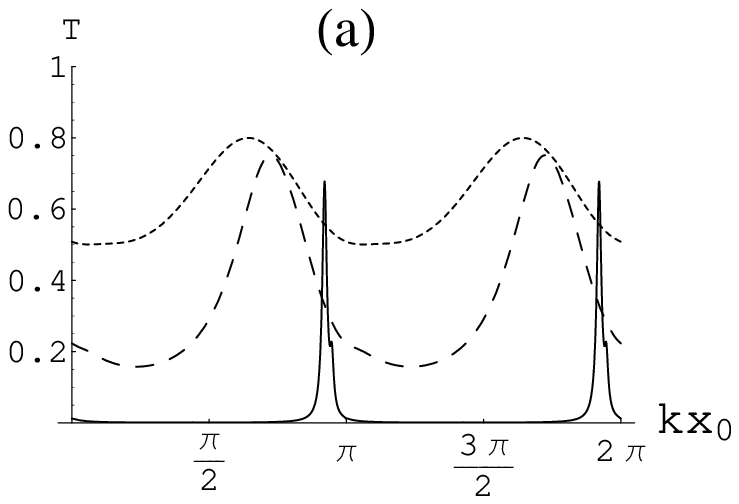}}
{\includegraphics[scale=0.7]{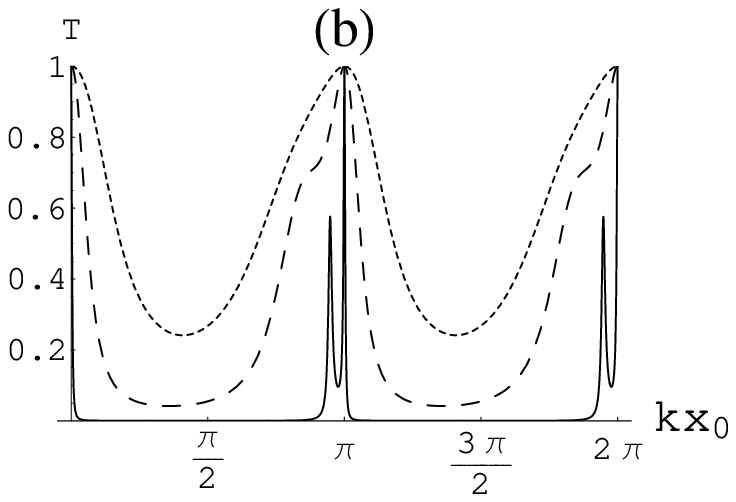}
              }}}
\caption{\footnotesize {Electron transmittivity $T$ as a function of
$kx_{0}$ when the electron is injected in the state $\ket{\s}$ with
the impurities prepared in the state $\ket{\Psi^{+}}$ (a) and
$\ket{\Psi^{-}}$ (b). Dotted, dashed and solid lines stand for
$\rho(E)J=1,2,10$, respectively.}}
\end{figure}

A striking behaviour however appears when the impurity spins are
prepared in the maximally entangled state $\ket{\Psi^{-}} =
(\ket{\s\g} -\ket{\g\s})/\sqrt{2}$: as shown in figure 3(b), the
wire becomes ``transparent" for $kx_{0}= n\pi$. In other words,
perfect transmittivity $T=1$ is reached at $kx_{0}= n\pi$ regardless
of the value of $\rho(E)J$, with peaks getting narrower for
increasing values of $\rho(E)J$. Furthermore, under the resonance
condition $kx_{0}= n\pi$, the spin state $\ket{\s}\ket{\Psi^{-}}$ is
transmitted unchanged. Note that this occurs even if
$\ket{\s}\ket{\Psi^-}$ belongs to the $s=1/2$ subspace where
spin-flip is allowed (see subsection \ref{subspace_s12}).
Importantly, this phenomenon takes place regardless of the electron
spin state. Indeed, in \ref{appendix} we demonstrate that the only
spin state $\ket{\chi}$ allowing perfect transparency of figure 3(b)
to occur is of the form
\begin{equation} \label {chi_transparent}
\ket{\chi}=\left(\alpha\ket{\s}+\beta\ket{\g}\right)\ket{\Psi^{-}}
\end{equation}
with arbitrary complex values of $\alpha$ and $\beta$.

\section{Conservation of $\mathbf{S}_{12}^2$} \label{conservation}

The effect of perfect transparency presented in the previous section
is clearly due to constructive quantum interference. In this section
we show how this phenomenon can be quantitatively analyzed in terms
of Hamiltonian symmetries. Let $\delta_{k}(x)$ and
$\delta_{k}(x-x_0)$ be the effective representations of $\delta(x)$
and $\delta(x-x_0)$, respectively, in a subspace of fixed energy
$E=\hbar^{2} k^2/2m^*$. Using the matrix representations of electron
orbital operators $\delta_{k}(x)$ and $\delta_{k}(x-x_0)$ in the
basis $\left\{ \ket{k},\ket{-k}\right\}$, it is straightforward to
prove that $\delta_{k}(x)=\delta_{k}(x-x_0)$ for $kx_{0}=n\pi$. When
this occurs the non-kinetic part $V$ of $H$ in equation (\ref{H})
assumes the effective representation
\begin{equation}\label{Vk}
V=J \, \mbox{\boldmath$\sigma$}\cdot \mathbf{S}_{12}
\,\delta_{k}(x)=\frac{J}{2}\left(\mathbf{S}^2-\bm{\sigma}^2-\mathbf{S}_{12}^2\right)\,\delta_{k}(x)
\end{equation}
with $\mathbf{S}_{12}=\mathbf{S}_{1}+\mathbf{S}_{2}$ being the total
spin of the two impurities. This means that for electron wave
vectors fulfilling the condition $kx_{0}=n\pi$, the operator
$\mathbf{S}_{12}^2$ (with quantum number $s_{12}$) becomes a
constant of motion whatever the strength of $J$. This is physically
reasonable since the condition $kx_{0}=n\pi$ implies that the
electron is found at $x=0$ and $x=x_0$ with equal probabilities and,
as a consequence, the two impurities are equally coupled to the
electron spin.

Furthermore, from equation (\ref{Vk}) it follows that $V$ vanishes
for $s=1/2$ and $s_{12}=0$. This is the case for the initial state
(\ref{chi_transparent}) as this is an eigenstate of $\mathbf{S}^{2}$
and $\mathbf{S}_{12}^{2}$ with quantum numbers $s=1/2$ and
$s_{12}=0$, respectively (see also \ref{appendix}). Therefore, when
this state is prepared and $kx_{0}=n\pi$, no spin-flip occurs and
the wire becomes transparent: an effective quenching of the
electron-impurities coupling takes place. This explains the results
of figure 3(b).

The same behaviour however does not occur for the state
$\ket{\s}\ket{\Psi^+}$ ($\ket{\g}\ket{\Psi^+}$) belonging to the
degenerate 2D eigenspace of $\mathbf{S}_{12}^{2}$ and $S_{z}$ with
quantum numbers $s_{12}=1$ and $m=1/2$ ($m=-1/2$), respectively.
Since the orthogonal state $\ket{\g}\ket{\s \s}$ ($\ket{\s}\ket{\g
\g}$) lies in the same eigenspace it follows that, when $kx_{0}=
n\pi$, the transmitted spin state will result in a linear
combination of $\ket{\s}\ket{\Psi^+}$ and $\ket{\g}\ket{\s \s}$
($\ket{\g}\ket{\Psi^+}$ and $\ket{\s}\ket{\g \g}$), implying
spin-flip and decoherence in agreement with typical behaviour of
magnetic impurities. This explains the decoherent behaviour of
figure 3(a).

\section{Entanglement controlled transmission and maximally entangled states QND scheme}
\label{ent_controlled}

The deeply different behaviours exhibited by $\ket{\Psi^-}$ and
$\ket{\Psi^+}$ at $kx_0=n\pi$ suggest that, in this regime, electron
transmission through the wire is strongly affected by the relative
phase $\varphi$ between the impurity spin states $\ket{\s\g}$ and
$\ket{\g\s}$ appearing in (\ref{family2}). In figure 4 we thus plot
the transmittivity $T$ when the electron is injected in an arbitrary
spin state $(\alpha\ket{\s}+\beta\ket{\g})$ with the impurities
prepared in a state (\ref{family2}) as a function of $\vartheta$ and
$\varphi$, for $kx_{0}= n\pi$ and $\rho(E) J=10$. Note how the
electron transmission indeed depends crucially on $\varphi$. The
maximum value of $T$ occurs when the impurities are prepared in the
singlet state $\ket{\Psi^{-}}$, while its minima occur for the
triplet state $\ket{\Psi^{+}}$.
\begin{figure}
 \includegraphics [scale=0.8]{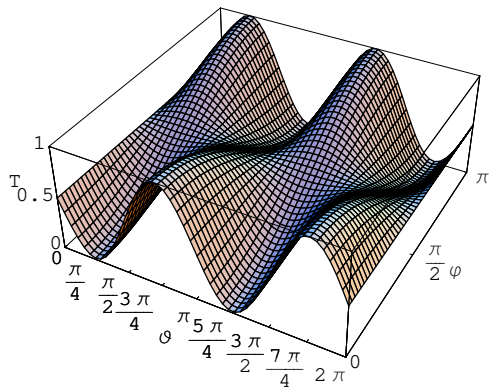}
 \caption{\label{teta_phi} Electron transmittivity $T$ at $kx_{0}=n\pi$ and $\rho(E)J=10$
 when the electron is injected in an arbitrary spin
state $(\alpha\ket{\s}+\beta\ket{\g})$ with the impurities prepared
in the state
$\cos{\vartheta}\ket{\s\g}+e^{i\varphi}\sin{\vartheta}\ket{\g\s}$.}
\end{figure}
In agreement with what was discussed in section \ref{conservation},
denoting by $T_{\Psi^{\pm}}$ the transmittivity for
$\ket{\Psi^{\pm}}$, decoherence effects cause $T_{\Psi^+}<1$ (it
gets smaller and smaller for increasing values of $\rho(E) J$) while
$T_{\Psi^{-}}=1$ due to occurrence of perfect transparency. To
explain why $T_{\Psi^-}$ and $T_{\Psi^+}$ are, respectively, maxima
and minima of $T$, we observe that the set of states (\ref{family2})
is spanned by $\ket{\Psi^{-}}$ and $\ket{\Psi^{+}}$. Since these two
states belong to orthogonal eigenspaces of the constant of motion
$\mathbf{S}_{12}^2$ a linear combination of them cannot exhibit
quantum interference effects and reduces to a statistical mixture.
This ensures that the transmittivity for a generic state
(\ref{family2}) will have intermediate values between $T_{\Psi^+}$
and $T_{\Psi^-}$.

The most remarkable result emerging from the above discussion is
that, within the set of initial impurity spins states
(\ref{family2}), maximally entangled states $\ket{\Psi^{-}}$ and
$\ket{\Psi^{+}}$ have the relevant property of maximizing or
minimizing electron transmission. We have chosen $\rho(E)J=10$ to
better highlight this behaviour, but this happens for any value of
$\rho(E)J$. This result suggests the appealing possibility to use
the relative phase $\varphi$ as a control parameter to modulate the
electron transmission in a 1D wire \cite{ciccarello}.

According to what was discussed in section \ref{conservation},
perfect transparency ensures that, once $\ket{\Psi^{-}}$ is set for
obtaining high conductivity of the device, this impurity spins state
will not be lost during transport of electrons through the wire.
However, the same is not true for the low conductivity-state
$\ket{\Psi^+}$ which is instead affected by spin-flip events and in
general will be destroyed by electron scattering (see section
\ref{conservation}). For the above entanglement
controlled-modulation to be correctly performed, it is thus required
that $\ket{\Psi^+}$ can be protected from spin-flip events. To
achieve this, conservation of $\mathbf{S}_{12}^2$ can again be
fruitfully used together with proper spin-filtering.

Assume thus that the electrons are injected in a fixed spin state,
let us say $\ket{\s}$. As discussed in section \ref{conservation},
in the regime $kx_0=n\pi$ conservation of $\mathbf{S}_{12}^2$
implies that $\ket{\s}\ket{\Psi^+}$ is transmitted (or reflected) as
a linear combination of $\ket{\s}\ket{\Psi^+}$ and
$\ket{\g}\ket{\s\s}$. It follows that if the electrons are analyzed
at the output of the wire in the same incoming spin state
$\ket{\s}$, the state $\ket{\Psi^+}$ of the impurity spins is
protected from spin-flip \cite{nota_spin_filtering}.

Let us denote by $T_{+}$ the spin-polarized transmission amplitude
that the electron is transmitted in the state $\ket{\s}$. In figures
5(a) and (b) we have plotted $T$ and $T_{+}$, respectively, for an
initial impurity spins state (\ref{family2}) with the electron
injected in the state $\ket{\s}$ and for $kx_0=n\pi$ and
$\rho(E)J=2$. Note how $T_{\Psi^+}$ turns out to be lowered in
figure 5(b) compared to 5(a). This is better visible in Fig. 5(c),
where  both $T$ and $T_{+}$, for $\varphi=0$, are plotted: in both
cases maxima and minima occur for $\ket{\Psi^-}$ and $\ket{\Psi^+}$,
respectively, but while maxima coincide, minima are lowered in the
spin-filtered case. Spin-filtering thus allows the entanglement
controlled-transmission task to work even more efficiently.
\begin{figure}[htbp]
\begin{center}
              {\hbox{ {\includegraphics[scale=0.6]{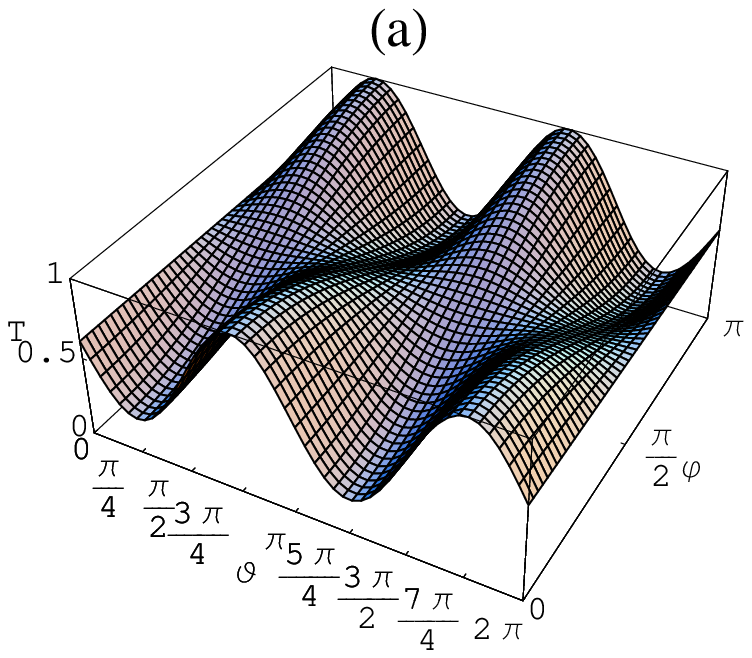}}
{\includegraphics[scale=0.6]{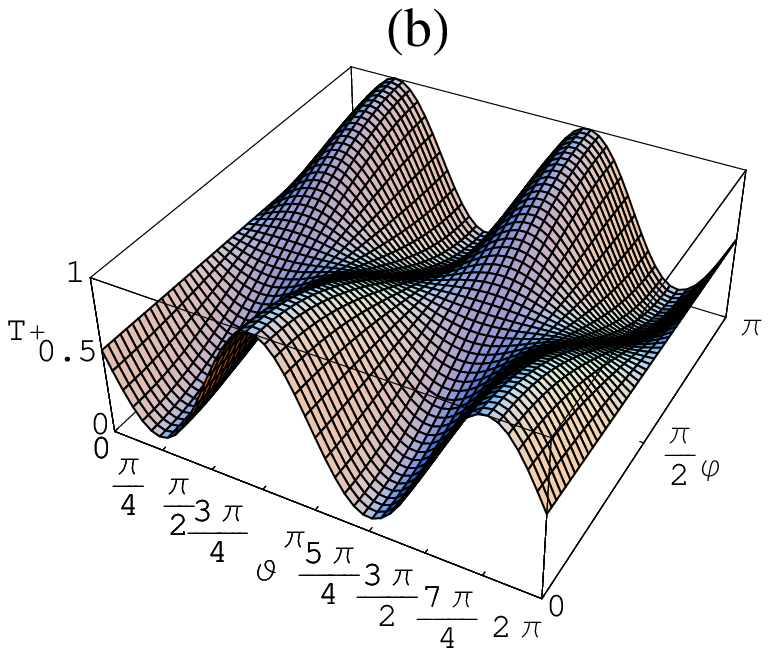}}{\includegraphics[scale=0.6]{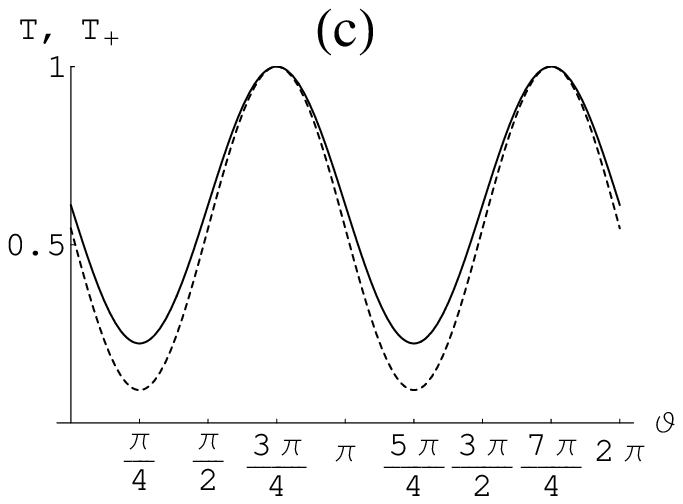}}}}
\caption{\footnotesize {Electron transmittivity $T$ (a) and
conditional electron transmittivity $T_+$ (b) at $kx_{0}=n\pi$ and
$\rho(E)J=2$ when the electron is injected in the state $\ket{\s}$
with the impurities prepared in the state
$\cos{\vartheta}\ket{\s\g}+e^{i\varphi}\sin{\vartheta}\ket{\g\s}$. A
comparison between $T$ (solid line) and $T_+$ (dashed line) for
$\varphi=0$ is detailed in Fig. 5(c).}}
\end{center}
\end{figure}
We have found that $T_{+}\simeq T$ for high values of $\rho(E)J$ as
in figure 3 for $\rho(E)J=10$. Thus in these cases no spin-filtering
is required to protect $\ket{\Psi^+}$.

Finally, we point out that the result showed in figure 4 opens the
possibility of a new maximally entangled states detection scheme.
Indeed, electron transmission can be regarded as a probe to detect
maximally entangled singlet and triplet states of two localized
spins within the family (\ref{family2}). In particular, it should be
clear from the above discussion that use of spin-filtering makes the
above setup  a quantum non-demolition (QND) detection scheme both
for $\ket{\Psi^-}$ and $\ket{\Psi^+}$. In particular, for the state
$\ket{\Psi^-}$, such scheme works as a QND even without
spin-filtering.

\section{Transmission properties for aligned impurity spins states}
\label{aligned_spins}

Not all the sets of maximally entangled states exhibit the effects
described in sections \ref{one_spin_up}, \ref{conservation} and
\ref{ent_controlled}. To show this, in this section, we consider a
different family of impurity spins states
\begin{equation} \label{uu_dd}
\ket{\phi (\vartheta , \varphi)} = \cos\vartheta\ket{\s\s} +
e^{i\varphi}\sin\vartheta\ket{\g\g}
\end{equation}
where again $\vartheta\in[0,2\pi]$ and $\varphi\in[0,\pi]$. Family
(\ref{uu_dd}) describes all the states in which the impurity spins
are aligned.

The transmittivity $T$ for an electron incoming in the up spin state
is shown in figures 6 (a), (b) and (c) for the cases $\vartheta = 0$
(a), $\vartheta = \pi /2$ (b) and $\vartheta = \pi /4$ with
arbitrary $\varphi$ (c). Note how the maximum of $T$ in the case of
figure 6(c) has an intermediate value between the maximum of $T$ of
figure 6(a) and the one of figure 6(b). Additionally, the results of
figure 6(c) do not depend on the value of $\varphi$. The above
behaviour can be easily understood once one realizes that, in the
case of the initial spin state (\ref{uu_dd}) the two impurities
indeed behave as if they were prepared in the mixed state
\begin{equation}
\rho=\cos^2 \vartheta \ket{\s \s} \langle {\s \s}|+\sin^2 \vartheta
\ket{\g \g} \langle {\g \g}|
\end{equation}
The phase $\varphi$ thus plays no role for the present family of
states and no  interference effect occurs. The reason for this is
that $\ket{\s}\ket{\s\s}$ and $\ket{\s}\ket{\g\g}$ are eigenstates
of the constant of motion $S_z$ with different quantum numbers
$m=3/2$ and $m=-1/2$, respectively. Therefore, unlike the set of
states (\ref{family2}), no quantum interference effects are
possible.

Additionally, note that while in the cases of figures 6 (b) and (c)
a loss of electron coherence is exhibited similarly to the cases of
figures 2(a), 2(b) and 3(a), a coherent behaviour completely
analogous to a FP interferometer with partially silvered mirrors
\cite{rossi} is observed when the impurities are prepared in the
state $\ket{\s\s}$ with the electron injected in the state
$\ket{\s}$, as illustrated in figure 6(a). Indeed, the spin state
$\ket{\s}\ket{\s\s}$ belongs to the non degenerate eigenspace
$s=3/2$, $m=3/2$ where spin-flip does not occur and the impurities
behave as if they were static (see subsection \ref{subspace_s32}).
However, we emphasize that at variance with perfect transparency
induced by the impurities' singlet state shown in figure 3(b), here
$T=1$ for values of $kx_0$ \emph{depending} on $\rho(E)J$ and only
when the electron spin is initially aligned with the spins of the
impurities.

The effect of transparency presented in figure 3(b) thus requires
neither the knowledge of the coupling constant $J$ nor any
constraint on the electron spin state to be observed.
\begin{figure}
              \includegraphics [scale=1]{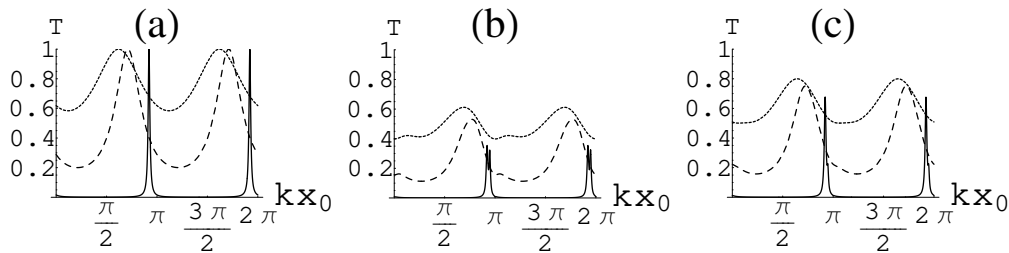}\caption{\footnotesize
{Electron transmittivity $T$ as a function of $kx_{0}$ when the
electron is injected in the state $\ket{\s}$ with the impurities in
the initial state $\ket{\s\s}$ (a), $\ket{\g\g}$ (b) and
$(\ket{\s\s} +e^{i\varphi}\ket{\g\g})/\sqrt{2}$ for arbitrary
$\varphi$ (c). Dotted, dashed and solid lines stand for
$\rho(E)J=1,2,10$, respectively. }} \label{fig1}
\end{figure}

\section{Generation of entangled states} \label{ent_gen}

To observe the entanglement dependent electron transmittivity
discussed in \ref{one_spin_up}, \ref{conservation} and
\ref{ent_controlled} one must be able to prepare the maximally
entangled states $\ket{\Psi^-}$ and $\ket{\Psi^+}$. In particular
this is required to observe the entanglement controlled
transmittivity illustrated in section \ref{ent_controlled}. Although
in our Hamiltonian model (\ref{H}) there is no direct interaction
between the two impurities, an indirect coupling via the electron
spin takes place, as it is implicit in the non-kinetic part of $H$.
In this section we thus show how electron-impurities scattering
itself can be used to generate the desired entanglement between the
impurity spins.

We first observe that it is enough to be able to generate only one
of the two states $\ket{\Psi^-}$ and $\ket{\Psi^+}$ since they can
be easily transformed into each other by simply introducing a
relative phase shift by means of a local field.  In the following,
we therefore show how the triplet state $\ket{\Psi^+}$ can be
generated by electron scattering. Generation of entangled states of
two magnetic impurities via electron scattering in 1D systems has
been recently investigated in \cite{yang, yasser, giorgi, nakazato}.
%A simplified scheme for entangling two distant magnetic impurities
%in a 1D wire via electron scattering has been recently proposed by
%Costa \emph{et al.} \cite{yasser}.
The basic idea is to inject an electron in the state $\ket{\s}$ with
the two impurities prepared in the state $\ket{\g\g}$. Due to
conservation of $S_z$ the transmitted spin state is of the form
\begin{equation}\label{ABC}
A \ket{\s}\ket{\g\g}+B\ket{\g}\ket{\s\g}+C\ket{\g}\ket{\g\s}
\end{equation}
It follows that if the transmitted electron is analyzed in the down
spin state $\ket{\g}$ the two impurities are projected in the
entangled state $B\ket{\s\g}+C\ket{\g\s}$ (apart from a
normalization factor) with probability $|B|^2+|C|^2$. For a fixed
electron energy, this state is not maximally entangled for any
strength of the electron-impurity coupling constant $J$
\cite{yasser}. However in \cite{yasser} the role played by the
distance $x_0$ between the impurities was not taken into account. In
the remaining of this section we will therefore consider an improved
entanglement generation scheme \cite{ciccarello} making use of the
exact knowledge of the energy eigenstates developed above. In
particular we will consider the regime $kx_0=n\pi$. In this case, in
addition to $S_z$, also $\mathbf{S}_{12}^2$ becomes a constant of
motion. It follows that the transmitted spin state will be of the
form (see also section \ref{conservation})
\begin{equation}\label{ABC2}
A' \ket{\s}\ket{\g\g}+B'\ket{\g}\ket{\Psi^+}
\end{equation}
An output filter selecting only transmitted electrons in the state
$\ket{\g}$ can thus be used to project the impurities into the state
$\ket{\Psi^{+}}$. Note that at variance with the analysis  discussed
in \cite{yasser}, this scheme ensures that a maximally entangled
state is always generated whatever the value of $J$. Furthermore, we
also know that this is the maximally entangled triplet state
$\ket{\Psi^+}$. The spin-polarized probability $T_{-}$ for the
electron to be transmitted in the state $\ket{\g}$ - that is to
project the impurities into $\ket{\Psi^+}$ - is plotted in figure 7
as a function of $\rho(E)J$.
\begin{figure}
 \includegraphics [scale=0.8]{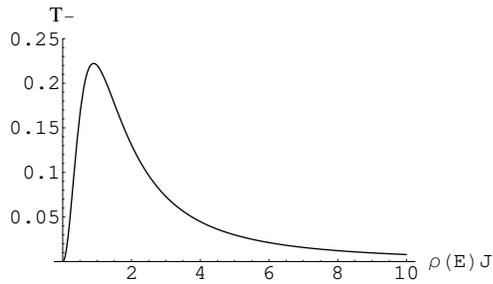}
 \caption{\label{P} Spin-polarized electron transmittivity $T_{-}$ at $kx_{0}=n\pi$ as a function of
 $\rho(E)J$ when the electron is injected in the state $\ket{\s}$ with the impurities prepared in the state
 $\ket{\g\g}$.}
\end{figure}
A probability larger than 20\% can be reached with $\rho(E)J\simeq
1$.

\section{Conclusions}

To discuss the possibility to observe the effects presented in this
paper, in particular the entanglement controlled transport, let us
assume an electron effective mass of $0.067\,m_0$ (as in GaAs
quantum wires) and two quantum dots -- each one of size 1 nm -- as
the impurities. Furthermore, the electron wavelength must be large
enough for the contact electron-dot potential of our Hamiltonian
model (\ref{H}) to be valid. This constraint implies that the
electron energy must not exceed 2 meV. In this case, requiring that
$\rho(E)J\simeq 1$ (i.e., as shown in section \ref{ent_gen}, the
optimal condition for generating entangled states of the impurities)
we obtain $J\simeq$ 1 eV{\AA}, which appears to be a reasonable
value of the electron-impurity coupling constant.

To prevent many-body effects, whose occurrence would make the single
electron-approach adopted in this work not valid, electrons could be
shot over an additional tunnel barrier before interacting with the
impurities as proposed in \cite{dalibard, schomerus}. This would
allow the injection of single electrons within a narrow energy range
well separated from the Fermi energy. Alternatively, this task could
be accomplished using a single-electron turnstile \cite{kouwenhoven}
as suggested e.g.\ in \cite{schomerus, jefferson}.

Finally we would like to comment on the effects of decoherence on
the interference phenomena described above. Some of the most
interesting features of electron spin in semiconductor
nanostructures are the long decoherence times, which is typically
larger than 100 nanoseconds (but can exceed in some cases the
microsecond) and the long coherence lengths, which can be longer
than 100 micrometers. Our approach is therefore able to predict an
observable effect. For instance for an electron energy of 2 meV, the
resonance condition $kx_0 =n\pi$ implies that $x_0$ must be in the
order of 100 nm. A coherence length of this order of magnitude is
common in a GaAs quantum wires at low temperatures (e.g.  see
\cite{Ferry}). Of course a stationary state approach to scattering
like the one we have used must be complemented with a dissipative
map for the evolution of the impurity spins when one is interested
in the steady state which can be obtained by the repeated injection
and detection of electrons in the wire.

In summary, in this paper we have considered a 1D wire with two
embedded spin-1/2 magnetic impurities. This system can be regarded
as the electron analogue of a Fabry Perot  interferometer in which
the two mirrors have internal spin degrees of freedom. Adopting an
appropriate quantum waveguide theory approach, we have derived all
the necessary transmission amplitudes at all orders in the
electron-impurity coupling constant. This has allowed us to
calculate the electron transmission properties for an arbitrary
initial spin state of the overall system. The typical behaviour is a
loss of electron coherence induced by spin-flip events due to
scattering by the magnetic impurities. However, when the maximally
entangled singlet state of the impurity spins is prepared, we have
found that perfect transparency of the wire is obtained regardless
of the electron spin state at wave vectors which do not depend on
the electron-impurity coupling constant. In the same regime, we have
found that electron transmittivity is maximized (minimized) by the
singlet (triplet) entangled states of the impurity spins. This
suggests a novel use of entanglement as a tool to modulate the
conductivity of a 1D wire. Additionally, the electron transmission
can be thought as a probe to detect maximally entangled singlet and
triplet states of two localized spins. When spin-filtering is
performed, this is a QND detection scheme for both these states,
while it works as a QND detection scheme for the singlet state even
without spin-filtering. These behaviours have been explained in
terms of the Hamiltonian symmetries, showing that appropriate
electron wave vectors allow for an effective conservation of the
squared total spin of the two impurities. Finally, we have proposed
a scheme to generate maximally entangled states via electron
scattering regardless of the electron-impurity coupling constant.

\section{Acknowledgments}

The authors thank the support from CNR (Italy) and GRICES
(Portugal). YO and VRV thank the support from Funda\c{c}\~{a}o para
a Ci\^{e}ncia e a Tecnologia (Portugal), namely through programs
POCTI/POCI and projects POCI/MAT/55796/2004 QuantLog and
POCTI-SFA-2-91, partially funded by FEDER (EU) as well as the
SQIG-IT EMSAQC initiative.

\appendix
\section{Set of perfectly ``transparent" states} \label{appendix}

In this Appendix we demonstrate that the only spin state
$\ket{\chi}$ allowing perfect transparency of figure 3(b) to occur
is of the form
$\ket{\chi}=\left(\alpha\ket{\s}+\beta\ket{\g}\right)\ket{\Psi^{-}}$.
To this we impose that the incoming and transmitted state coincide
for values of $kx_0$ which do not depend on $\rho(E)J$ . This yields
the conditions
\begin{equation} \label{trasparency}
\langle s_{e2};s,m \ket{\chi}=\gamma_{s_{e2},s,m}(\chi)
\end{equation}
Let us define the following matrix of transmission amplitudes in the subspace $s=1/2$
\begin{eqnarray}
\textbf{t}= \left(\begin{array}{cc}
    t^{(0;1/2)}_{0}  &    t^{(1;1/2)}_{0} \\
    t^{(0;1/2)}_{1} &    t^{(1;1/2)}_{1}
  \end{array} \right )
\end{eqnarray}
Taking into account (\ref{alpha_coeff}) and the selection rules for
$s_{e2}$ and $s_{e2}'$, conditions (\ref{trasparency}) reduce to an
equation for $s=3/2$ and a homogenous linear system of two equations
for $s=1/2$
\begin{eqnarray}
\left(t^{(1;3/2)}_{1} -1\right) \langle 1;3/2,m \ket{\chi}&=&0  \label{transp_s32} \\
\left(\textbf{t}-\textbf{I}\right)
  \times
  \left(\begin{array}{cc}
    \langle 0;1/2,m \ket{\chi} \\
    \langle 1;1/2,m \ket{\chi}
  \end{array} \right )
  &=&0  \label{transp_s12}
\end{eqnarray}
where $\textbf{I}$ is the $2\times 2$ identity matrix.

Let us assume that $\ket{\chi}$ has a non vanishing projection on
$s=3/2$. This means that $\langle 1;3/2,m \ket{\chi}\neq 0$ for at
least one $m$ and condition (\ref{transp_s32}) gives
$t^{(1;3/2)}_{1}=1$ which implies $|t^{(1;3/2)}_{1}|^2=1$.
$|t^{(1;3/2)}_{1}|^2$ is the transmittivity of a wire with two
static impurities of potential $J/4$ (see subsection
\ref{subspace_s32}) which is completely analogous to a FP
interferometer with partially reflecting mirrors. Since in this case
a resonance condition (transmittivity 1) occurs at values of $kx_0$
\emph{depending} on the mirror reflectivity \cite{rossi}, we
conclude that $t^{(1;3/2)}_{1}=1$ cannot occur at values of $kx_0$
independent on $\rho(E)J$ as for the case of figure 3(b).

It follows that $\langle 1;3/2,m \ket{\chi}=0\,\,\forall \,\,m$ (remind that coefficients $t_{s_{e2}}^{(s_{e2}';s)}$ are
$m$-independent). This implies that the spin states allowing occurrence of perfect transparency must fully lie in the subspace
$s=1/2$. Such states can be determined by requiring that linear system (\ref{transp_s12}) has non trivial solutions, that is
\begin{equation}\label{det_null}
\det{\left(\textbf{t-I}\right)}=0
\end{equation}
which with the help of (\ref{t0t1_1}) and (\ref{t0t1_2}) is
explicitly written as
\begin{equation}\label{det_null2}
\frac{3}{\delta} (e^{2ikx_0}-1)  (\pi \rho(E)J)^3 \left[3\pi
\rho(E)J\,(e^{2ikx_0}-1)+32i \right]=0
\end{equation}
with $\delta$ given by (\ref{delta}). Since the factor in square
brackets cannot vanish for real $kx_0$ equation (\ref{det_null2}) is
fulfilled for
\begin{equation} \label{kx0_eq_npi}
kx_0=n\pi
\end{equation}
with $n$ integer. Replacement of (\ref{kx0_eq_npi}) in
(\ref{transp_s12}) yields the $\rho(E)J$-independent solution
\begin{equation} \label{solution}
\langle 0;1/2,m \ket{\chi}=\frac{\langle 1;1/2,m
\ket{\chi}}{\sqrt{3}}
\end{equation}
with arbitrary coefficients $\langle 0;1/2,m \ket{\chi}$
($m=-1/2,1/2$). Rewriting these as $\langle 0;1/2,1/2
\ket{\chi}=\alpha/2$ and $\langle 0;1/2,-1/2 \ket{\chi}=\beta/2$,
$\ket{\chi}$ turns out to be of the form
\begin{eqnarray} \label{chi_sviluppo}
\fl\ket{\chi}=\alpha \left( \frac{1}{2}\ket{0;1/2,1/2} +\frac{\sqrt{3}}{2}\ket{1;1/2,1/2} \right)+\beta \left(
\frac{1}{2}\ket{0;1/2,-1/2}+\frac{\sqrt{3}}{2}\ket{1;1/2,-1/2} \right)\nonumber\\
\end{eqnarray}
Using Clebsh-Gordan coefficients, spin states inside brackets turn
out to be
\begin{eqnarray}\label{psi_meno_up}
\frac{1}{2}\ket{0;1/2,1/2}
+\frac{\sqrt{3}}{2}\ket{1;1/2,1/2}&=&\ket{\s}\ket{\Psi^-} \\
\frac{1}{2}\ket{0;1/2,-1/2} +\frac{\sqrt{3}}{2}\ket{1;1/2,-1/2}&=&\ket{\g}\ket{\Psi^-} \label{psi_meno_down}
\end{eqnarray}
Substitution of (\ref{psi_meno_up}) and (\ref{psi_meno_down}) into (\ref{chi_sviluppo}) proves that spin states exhibiting perfect
transparency are of the form $ \ket{\chi}=\left(\alpha\ket{\s}+\beta\ket{\g}\right)\ket{\Psi^{-}}$.


\begin{thebibliography}{20}
\bibitem{nota1} Of course, we could also choose the scheme in which
the electron is first coupled to impurity 1, while the scheme in
which the two impurities are first coupled together is not
convenient for the present problem.
\bibitem{nota_spin_filtering} We emphasize again that conservation of $\mathbf{S}_{12}^2$
is crucial for this spin-flip protection scheme to work. Indeed, if
$kx_0\neq n\pi$ the same spin-filtering procedure would leave the
impurity spins in a linear combination of $\ket{\s\g}$ and
$\ket{\g\s}$, this state being in general different from
$\ket{\Psi^+}$.
\end{thebibliography}
\end{document}